\documentstyle[12pt]{article}

\input{tcilatex}

\begin{document}

\title{A General $SU(2)$ Formulation for Quantum Searching with Certainty}
\author{Jin-Yuan Hsieh$^{1}$, Che-Ming Li$^{2}$ \\
$^{1}$Department of Mechanical Engineering, Ming Hsin Institute \\
of Technology, Hsinchu 30441, Taiwan.\\
$^{2}$Institute and Department of Electrophysics,National Chiao\\
Tung University, Hsinchu 30050, Taiwan.}
\date{}
\maketitle

\begin{abstract}
A general quantum search algorithm with arbitrary unitary transformations
and an arbitrary initial state is considered in this work. To serach a
marked state with certainty, we have derived, using an $SU(2)$
representation: (1) the matching condition relating the phase rotations in
the algorithm, (2) a concise formula for evaluating the required number of
iterations for the search, and (3) the final state after the search, with a
phase angle in its amplitude of unity modulus. Moreover, the optimal choices
and modifications of the phase angles in the Grover kernel is also studied.
\end{abstract}

Quantum mechanical algorithms have recently become very popular in the field
of computation science because they can speed up a computation over
classical algorithms. Famous examples include the factorizing algorithm
discovered by Shor\cite{shor} and the quantum search algorithm
well-develpoed by Grover\cite{grover1}\cite{grover2}. The latter is what we
intend to deal with in this work. If there is an unsorted database
containing $N$ items, and out of which only one marked item satisfies a
given condition, then using Grover's algorithm one will find the object in $%
O(\sqrt{N})$ quantum mechanical steps instead of $O(N)$ classical steps. It
has been shown that Grover$^{^{\prime }}$s original algorithm is optimal %
\cite{optimal1}\cite{optimal2}\cite{optimal3}. But Grover's \ algorithm
provides a high probability in finding the object only for a large $N$. The
probability will be lower as $N$ decreases. Grover\cite{grover3}, however,
also proposed that the Walsh-Hadamard transformation used in the original
version can be replaced by almost any arbitrary unitary operator and the
phase angles of rotation can be arbitrarily used as well, instead of the
original $\pi $-angles. The utility of the arbitrary phase angles in fact
can provide the possibility for finding the marked item with certainty, no
matter whether $N$ \ is large or not, if these angles obey a so-called
matching condition.

Some typical literatures concerning with the matching condition will be
mentioned here. Long et al.\cite{long1} \cite{long2}have derived the
relation $\phi =\theta $, where $\phi $ and $\theta $ are the phases used in
the algorithm, using an $SO(3)$ picture. H\o yer\cite{hoyer} , on the other
hand, has proved a relation $\tan (\phi /2)=\tan (\theta /2)(1-2/N)$, and
claimed that the relation $\phi =\theta $ is an approximation to this case.
Recently, a more general matching condition has been derived by Long et al. %
\cite{long3} , also using the $SO(3)$ picture . In the last article,
however, only the certainty for finding the marked state is ensured. In fact
a phase angle appearing in the amplitude of the final state after searching
will remain. If the final state should be necessary for a future
application, i.e., if it should interact with other states, this phase angle
will be important for quantum interferences, but it can not be given in the $%
SO(3)$ representation. We therefore intend to derive the matching condition
in the $SU(2)$ picture. Besides, we will also give a more concise formula
for evaluating the number of the iterations needed in the searching and
deduce the final state in a complete form as $e^{i\delta }\left| \tau
\right\rangle $, where $\left| \tau \right\rangle $ is the marked state. The
optimal choice of the phase angles will be discussed, too.

Suppose in a two-dimensional, complex Hilbert space we have a marked state $%
\left| \tau \right\rangle $ to be searched by successively operating a
Grover's kernel $G$ on an arbitrary initial state $\left| s\right\rangle $.
The Grover kernel is a product of two unitary operators $G_{\tau }$ and $%
G_{\eta }$, given by%
\begin{eqnarray}
G_{\tau } &=&I+(e^{i\phi }-1)\left| \tau \right\rangle \left\langle \tau
\right| \text{,} \\
G_{\eta } &=&I+(e^{i\theta }-1)U\left| \eta \right\rangle \left\langle \eta
\right| U^{-1}\text{ ,}  \nonumber
\end{eqnarray}%
where $U$ is an arbitrary unitary operator, $\left| \eta \right\rangle $ is
another unit vector in the space, and $\phi $ and $\theta $ are two phase
angles. It should be noted that the phases $\phi $ and $\theta $ actually
are the differences $\phi =\phi _{2}-\phi _{1}$ and $\theta =\theta
_{2}-\theta _{1}$, where $\phi _{2}$, $\phi _{1}$, $\theta _{2}$, and $%
\theta _{1}$, as depicted in refs.\cite{galindo} \cite{li} , denote the
rotating angles to $\left| \tau \right\rangle $, the vector orthogonal to $%
\left| \tau \right\rangle $, $U\left| \eta \right\rangle $, and the vector
orthogonal to $U\left| \eta \right\rangle $, respectively. The Grover kernel
can be expressed in a matrix form as long as an orthonormal set of basis
vectors is designated, so we simply choose%
\begin{equation}
\left| I\right\rangle =\left| \tau \right\rangle \text{ and }\left|
II\right\rangle =(U\left| \eta \right\rangle -U_{\tau \eta }\left| \tau
\right\rangle )/l\text{ ,}
\end{equation}%
where $U_{\tau \eta }=\left\langle \tau \right| U\left| \eta \right\rangle $
and $l=(1-\left| U_{\tau \eta }\right| ^{2})^{1/2}$. Letting $U_{\tau \eta
}=\sin (\beta )e^{i\alpha }$, we can write, from (2),

\begin{equation}
U\left| \eta \right\rangle =\sin (\beta )e^{i\alpha }\left| I\right\rangle
+\cos (\beta )\left| II\right\rangle \text{ ,}
\end{equation}%
and the Grover kernel can now be written%
\begin{eqnarray}
G &=&-\text{ }G_{\eta }G_{\tau }  \nonumber \\
&=&-\left[ 
\begin{array}{cc}
e^{i\phi }(1+(e^{i\theta }-1)\sin ^{2}(\beta )) & (e^{i\theta }-1)\sin
(\beta )\cos (\beta )e^{i\alpha } \\ 
e^{i\phi }(e^{i\theta }-1)\sin (\beta )\cos (\beta )e^{-i\alpha } & 
1+(e^{i\theta }-1)\cos ^{2}(\beta )%
\end{array}%
\right] \text{.}
\end{eqnarray}

In the searching process, the Grover kernel is successively operated on the
initial state $\left| s\right\rangle $. We wish that after, say, $m$
iterations the operation the final state will be orthogonal to the basis
vector $\left| II\right\rangle $ so that the probability for finding the
marked state $\left| \tau \right\rangle $ will exactly be unity.
Alternatively, in mathematical expression, we wish to fulfill the requirement

\begin{equation}
\left\langle II\right| G^{m}\left| s\right\rangle =0\text{ ,}
\end{equation}%
since then

\begin{equation}
\left| \left\langle \tau \right| G^{m}\left| s\right\rangle \right| =\left|
\left\langle I\right| G^{m}\left| s\right\rangle \right| =1\text{ .}
\end{equation}

The eigenvalues of the Grover kernel $G$ are

\begin{equation}
\lambda _{1,2}=-e^{i(\frac{\phi +\theta }{2}\pm w)}\text{ ,}
\end{equation}%
where the angle $w$ is defined by

\begin{equation}
\cos (w)=\cos (\frac{\phi -\theta }{2})-2\sin (\frac{\phi }{2})\sin (\frac{%
\theta }{2})\sin ^{2}(\beta )\text{ .}
\end{equation}%
The normalized eigenvectors associated with these eigenvalues are computed:

\begin{equation}
\left| g_{1}\right\rangle =\left[ 
\begin{array}{c}
e^{-i\frac{\phi }{2}}e^{i\alpha }\cos (x) \\ 
\sin (x)%
\end{array}%
\right] \text{ \ ,}\left| g_{2}\right\rangle =\left[ 
\begin{array}{c}
-\sin (x) \\ 
e^{i\frac{\phi }{2}}e^{-i\alpha }\cos (x)%
\end{array}%
\right] \text{.}
\end{equation}%
In expression (9), the angle $x$ is defined by

\[
\sin (x)=\sin (\frac{\theta }{2})\sin (2\beta )/\sqrt{l_{m}}\text{,} 
\]%
where%
\begin{eqnarray*}
l_{m} &=&(\sin (w)+\sin (\frac{\phi -\theta }{2})+2\cos (\frac{\phi }{2}%
)\sin (\frac{\theta }{2})\sin ^{2}(\beta ))^{2}+(\sin (\frac{\theta }{2}%
)\sin (2\beta ))^{2} \\
&=&2\sin (w)(\sin (w)+\sin (\frac{\phi -\theta }{2})+2\cos (\frac{\phi }{2}%
)\sin (\frac{\theta }{2})\sin ^{2}(\beta ))\text{.}
\end{eqnarray*}%
The matrix $G^{m}$ can be simply expressed by $G^{m}=\lambda _{1}^{m}\left|
g_{1}\right\rangle \left\langle g_{1}\right| +\lambda _{2}^{m}\left|
g_{2}\right\rangle \left\langle g_{2}\right| $, so we have

\begin{equation}
G^{m}=(-1)^{m}e^{im(\frac{\phi +\theta }{2})}\left[ 
\begin{array}{cc}
e^{imw}\cos ^{2}(x)+e^{-imw}\sin ^{2}(x) & e^{-i\frac{\phi }{2}}e^{i\alpha
}i\sin (mw)\sin (2x) \\ 
e^{i\frac{\phi }{2}}e^{-i\alpha }i\sin (mw)\sin (2x) & e^{imw}\sin
^{2}(x)+e^{-imw}\cos ^{2}(x)%
\end{array}%
\right] \text{.}
\end{equation}

The initial state $\left| s\right\rangle $ in this work is considered to be
an arbitrary unit vector in the space and is given by

\begin{equation}
\left| s\right\rangle =\sin (\beta _{0})\left| I\right\rangle +\cos (\beta
_{0})e^{iu}\left| II\right\rangle \text{.}
\end{equation}%
The requirement (5) implies that both the real and imagine parts of the term 
$\left\langle II\right| G^{m}\left| s\right\rangle $ are zero, so, as
substituting (10) and (11) into (5), one will eventually obtain the two
equations:

\begin{equation}
-\sin (mw)\sin (\frac{\phi }{2}-\alpha -u)\sin (2x)\sin (\beta _{0})+\cos
(mw)\cos (\beta _{0})=0\text{,}
\end{equation}

\begin{equation}
\sin (mw)\cos (\frac{\phi }{2}-\alpha -u)\sin (2x)\sin (\beta _{0})-\sin
(mw)\cos (2x)\cos (\beta _{0})=0\text{.}
\end{equation}%
Equation (13), by the definition of the angle $x$, will reduce to the
matching condition

\begin{equation}
(\sin (\frac{\phi -\theta }{2})+2\cos (\frac{\phi }{2})\sin (\frac{\theta }{2%
})\sin ^{2}(\beta ))\cos (\beta _{0})=\sin (\frac{\theta }{2})\sin (2\beta
)\cos (\frac{\phi }{2}-\alpha -u)\sin (\beta _{0})\text{,}
\end{equation}%
which is identical to the relation derived by Long et al.\cite{long3}:

\begin{equation}
\tan (\frac{\phi }{2})=\tan (\frac{\theta }{2})(\frac{\cos (2\beta )+\sin
(2\beta )\tan (\beta _{0})\cos (\alpha +u)}{1-\tan (\beta _{0})\tan (\frac{%
\theta }{2})\sin (2\beta )\sin (\alpha +u)})\text{.}
\end{equation}%
Equation (12), under the satisfaction of the matching condition (14), or
(15), will reduce to a concise formula for evaluating the number of
iterations $m$:

\begin{equation}
\cos (mw+\sin ^{-1}(\sin (\beta _{0})\sin (\frac{\phi }{2}-\alpha -u)))=0%
\text{.}
\end{equation}%
By equation (16), one can compute the number $m$

\begin{equation}
m=\left\lfloor f\right\rfloor \text{,}
\end{equation}%
where $\lfloor $ $\rfloor $ denotes the smallest integer greater than the
quantity in it, and the function $f$ is given by

\begin{equation}
f=\frac{\frac{\pi }{2}-\sin ^{-1}(\sin (\beta _{0})\sin (\frac{\phi }{2}%
-\alpha -u))}{\cos ^{-1}(\cos (\frac{\phi -\theta }{2})-2\sin (\frac{\phi }{2%
})\sin (\frac{\theta }{2})\sin ^{2}(\beta ))}\text{.}
\end{equation}%
It can also be shown that if the matching condition is fulfilled, then after 
$m$ searching iterations the final state will be

\begin{equation}
G^{m}\left| s\right\rangle =e^{i\delta }\left| \tau \right\rangle =e^{i\left[
m(\pi +\frac{\phi +\theta }{2})+\Omega \right] }\left| \tau \right\rangle 
\text{,}
\end{equation}%
where the angle $\Omega $ is defined by

\begin{equation}
\Omega =\tan ^{-1}(\cot (\frac{\phi }{2}-\alpha -u))\text{.}
\end{equation}%
The phase angle appearing in the amplitude of the final state will be
important for quantum interferences if possibly the state should interact
with other states in a future application, so we would had better remain it
as the present form.

The matching condition (14), or (15), relates the angles $\phi $, $\theta $, 
$\beta $, $\beta _{0}$ , and $\alpha +u$ for finding a marked state with
certainty. If $\beta $, $\beta _{0}$ and $\alpha +u$ are designated, then $%
\phi =\phi (\theta )$ is deduced by the matching condition. As $\phi (\theta
)$ is determined, we then can evaluate by (18) the value of $f=f(\phi
(\theta ),$ $\theta )$ and consequently decide by (17) the number of
iterations $m$. The functions $\phi (\theta )$ and $f(\theta )$ for some
particular designations of $\beta $, $\beta _{0}$ and $\alpha +u$ have been
shown in Figs.1 and 2. These examples have schematically depicted that
theoretically we can establish a tabulated chart of possible choices between
all of the phases for finding a marked state with certainty. It is worth
noticing that as $\alpha +u=0$ and $\beta =\beta _{0}$, the matching
condition recovers $\phi =\theta $ automatically since then eq. (13) becomes
an identity, and accordingly one has

\begin{equation}
f=\frac{\frac{\pi }{2}-\sin ^{-1}(\sin (\frac{\phi }{2})\sin (\beta ))}{%
2\sin ^{-1}(\sin (\frac{\phi }{2})\sin (\beta ))}\text{, for }\phi =\theta 
\text{.}
\end{equation}%
This is the case discussed in ref.\cite{long2}; an example can be read by
the straight line of unity slope for $\beta $=$\beta _{0}$=$10^{-4}$ and the
corresponding $f$ vs $\theta $ variation in Fig.1. It can also be shown that
the matching condition (14) will recover the relation considered by H\o yer%
\cite{hoyer}:

\begin{equation}
\tan (\frac{\phi }{2})=\tan (\frac{\theta }{2})\cos (2\beta )\text{, for }%
\cos (\phi /2-\alpha -u)=0\text{.}
\end{equation}%
In Figs.1 and 2 we have shown by the cross marks some particular examples of
this special case.

Observing Figs 1 and 2, one realizes that for every designation of $\beta $, 
$\beta _{0}$ and $\alpha +u$, the optimal choices for $\phi $ and $\theta $
is letting $\phi =\theta =\pi $, since then the corresponding $f$ is minimum
under the fact $df/d\theta =(\partial f/\partial \phi )(d\phi /d\theta
)+\partial f/\partial \theta =0$, for $\phi =\theta =\pi $. We thus denote
the optimal value of $m$ by

\begin{equation}
m_{op}=\left\lfloor \min (f)\right\rfloor =\left\lfloor \frac{\frac{\pi }{2}%
-\sin ^{-1}(\sin (\beta _{0})\cos (\alpha +u))}{2\beta }\right\rfloor \text{.%
}
\end{equation}%
With the choice of $m_{op}$, however, one need to modify the phases $\theta $
and $\phi (\theta )$ to depart from $\pi $ so that the matching condition is
satisfied again. For example, if $\alpha +u=0$, $\beta _{0}=10^{-4}$ and $%
\beta =0.7$ are designated, then the minimum value of $f$ will be $\min
(f)=0.56$ . So we choose $m_{op}=1$ and the modified phases are $\theta
_{op}=(1\pm 0.490)\pi $ and $\phi _{op}=(1\pm 0.889)\pi $ ,respectively.
This example has been shown by the marked entire circles in Fig.1. It is
worth noticing again that under the choice of $m_{op}$ the modified $\phi $
and $\theta $ for the special case considered by Long\cite{long2} will be

\[
\phi _{op}=\theta _{op}=\left\lfloor \min (f)\right\rfloor =2\sin ^{-1}(%
\frac{\sin (\frac{\pi }{4m_{op}+2})}{\sin (\beta )})\text{,} 
\]%
where

\[
m_{op}=\left\lfloor \frac{\frac{\pi }{2}-\beta }{2\beta }\right\rfloor \text{%
.} 
\]%
This is in fact a special case in which the phases $\phi _{op}$ and $\theta
_{op}$ can be given by a closed-form formula.

To summarize, using the $SU(2)$ representation, we have derived the matching
condition (14) for finding with certainty a marked state with arbitrary
unitary transformations and an arbitrary initial state. The formula (17),
together with (18), has also been deduced for evaluating the required number
of interations for the search. Moreover, the final state with a phase angle
in its amplitude, which can not be given by the $SO(3)$ picture used in ref.%
\cite{long3}, has consequently obtained. The optimal choice $\phi =\theta
=\pi $ under any designation of $\beta $, $\beta _{0}$ and $\alpha +u$ has
been shown. However, for finding the marked state with certainty, the phases 
$\phi $ and $\theta $ need to be modified since that $m_{op}$ must be an
integer. An example to depict the modification of $\phi $ and $\theta $,
therefore, has also been given.

\newpage \pagebreak \pagebreak

\FRAME{ftbpFU}{2.7622in}{3.378in}{0pt}{\Qcb{Variations of $\protect\phi (%
\protect\theta )$ (solid) and $f(\protect\theta )$ (broken), for $\protect%
\alpha +u=0$, $\protect\beta _{0}=10^{-4}$, and $\protect\beta =10^{-4}$
(1), $10^{-2}$ (2), $0.5$ (3) and $0.7$ (4), respectively. The cross marks
denote the special case of H\o yer[10], while the entire cirles correspond
to the optimal choices of $\protect\phi _{op}$ and $\protect\theta _{op}$
for $\protect\alpha +u=0$, $\protect\beta _{0}=10^{-4}$ and $\protect\beta %
=0.7$. The solid straight line 1 corresponds the case $\protect\phi =\protect%
\theta $, while the solid curve 2 is only approximately close to the former.}%
}{}{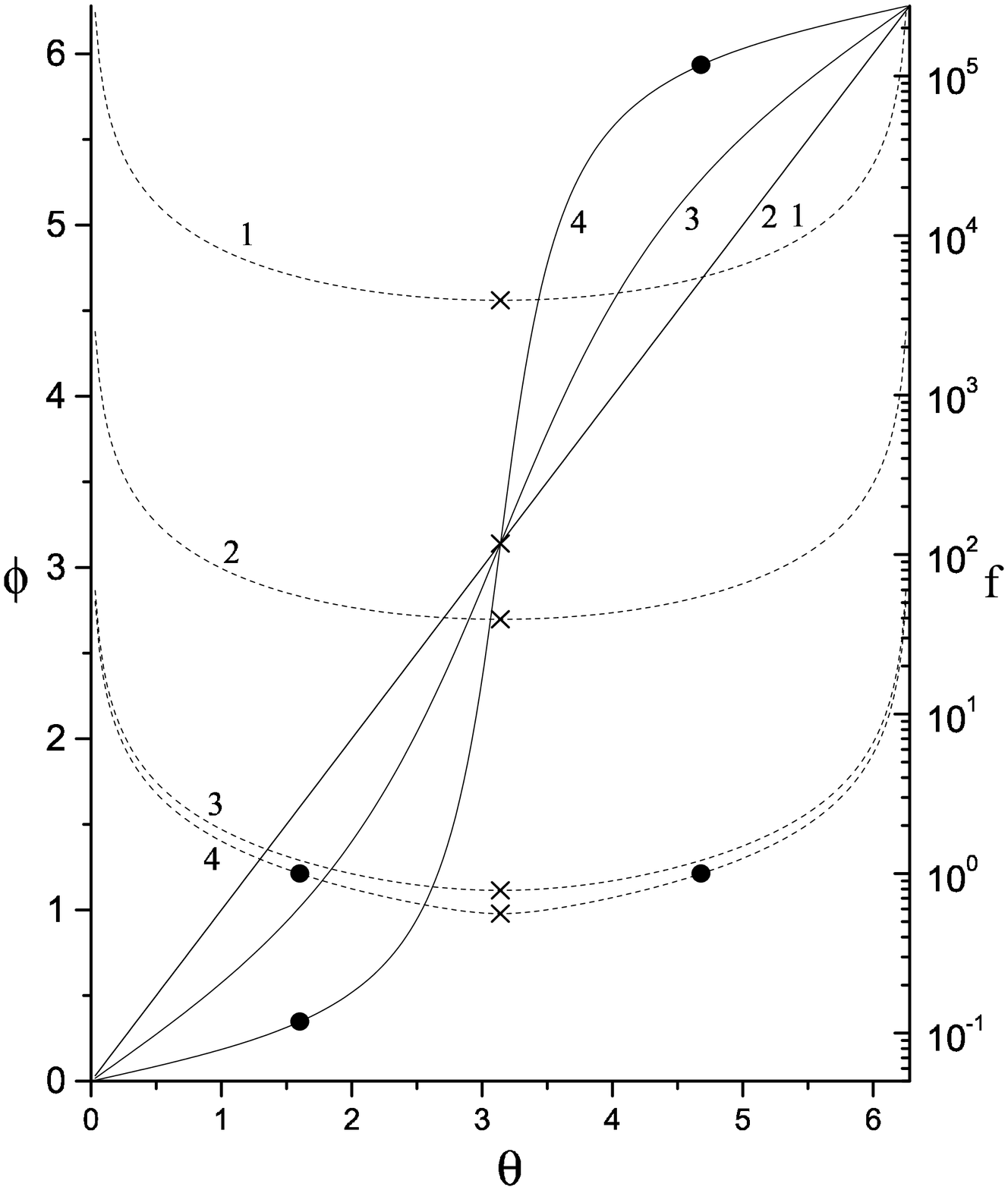}{\special{language "Scientific Word";type
"GRAPHIC";maintain-aspect-ratio TRUE;display "USEDEF";valid_file "F";width
2.7622in;height 3.378in;depth 0pt;original-width 8.2313in;original-height
10.0854in;cropleft "0.022857";croptop "1.083968";cropright
"1.022857";cropbottom "0.083968";filename '../PRA/fig1.eps';file-properties
"XNPEU";}}

\FRAME{ftbpFU}{2.7475in}{3.378in}{0pt}{\Qcb{Variations of $\protect\phi (%
\protect\theta )$ (solid) and $f(\protect\theta )$ (broken), for $\protect%
\alpha +u=0.1$, $\protect\beta _{0}=0.1$, and $\protect\beta =10^{-4}$ (1), $%
10^{-2}$ (2), $0.5$ (3) and $0.7$ (4), respectively. The cross marks denote
the special case of H\o ye[10]. The solid curves 1 and 2 are very close, and
both of them are only approximately close to the line $\protect\phi =\protect%
\theta $.}}{}{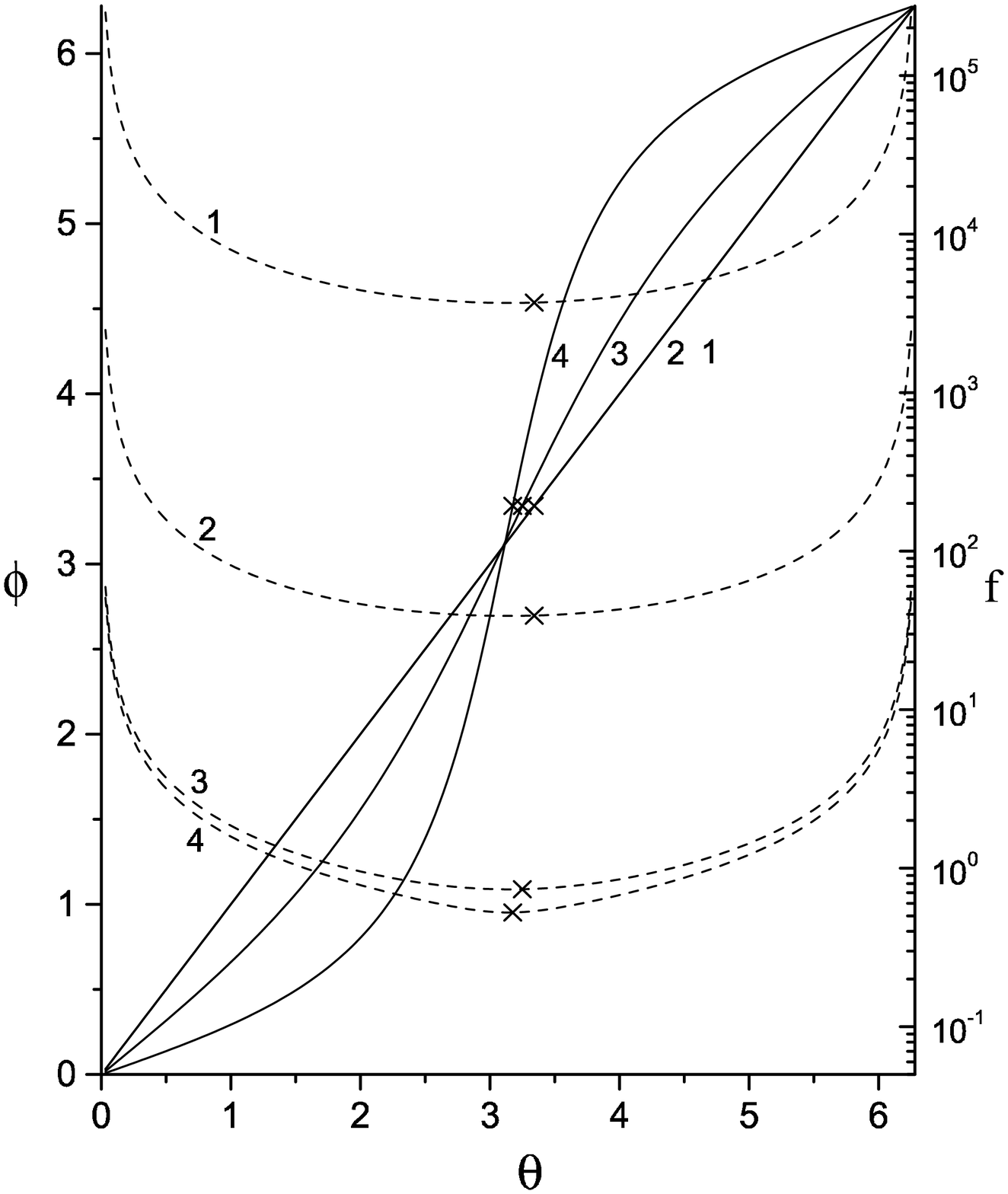}{\special{language "Scientific Word";type
"GRAPHIC";maintain-aspect-ratio TRUE;display "USEDEF";valid_file "F";width
2.7475in;height 3.378in;depth 0pt;original-width 8.188in;original-height
10.0854in;cropleft "0.026808";croptop "1.068405";cropright
"1.026808";cropbottom "0.068405";filename '../PRA/fig2.eps';file-properties
"XNPEU";}}

\end{document}